\documentclass[twocolumn,secnumarabic,amssymb,superscriptaddress,nobibnotes,aps,pre,10pt]{revtex4-1}
\usepackage[toc,page,title,titletoc,header]{appendix}
\usepackage{stmaryrd}
\usepackage{amsmath,amsthm}
\usepackage{mathrsfs}
\usepackage{tabularx}
\usepackage{subfigure}
\usepackage{graphicx}
\usepackage{epsfig}
\usepackage{color}

\newcommand{\bmath}[1]{\mbox{\boldmath{$#1$}}}
\renewcommand{\vec}[1]{\mathbf{#1}}
\def\subFV{\scriptscriptstyle{FV}}
\def\subFS{\scriptscriptstyle{FS}}
\def\subVS{\scriptscriptstyle{VS}}
\def\Gammasub{\hat{\Gamma}}
\def\Tsub{\hat{\bmath{\tau}}}
\def\Nsub{\hat{\mathbf{n}}}

\begin{document}

\title{Solid-state dewetting on curved substrates}

\author{Wei Jiang}
\email{Corresponding author: jiangwei1007@whu.edu.cn}
\affiliation{School of Mathematics and Statistics,
Wuhan University, Wuhan 430072, China}

\author{Yan Wang}
\affiliation{Beijing Computational Science Research Center,
Beijing 100193, P.R. China}

\author{David J. Srolovitz}
\affiliation{Departments of Materials Science and Engineering {\rm \&} Mechanical Engineering
and Applied Mechanics, University of Pennsylvania, Philadelphia, PA 19104, USA}

\author{Weizhu Bao}
\affiliation{Department of Mathematics, National University of
Singapore, Singapore, 119076}


\begin{abstract}

Based on the thermodynamic variation to the free energy functional, we propose a sharp-interface model for simulating solid-state dewetting of thin films on rigid curved substrates in two dimensions. This model describes the interface evolution which occurs through surface diffusion-controlled mass transport and contact point migration along the curved substrate. Furthermore, the surface energy anisotropy is easily included into the model, and the contact point migration is explicitly described by the relaxed contact angle boundary condition. We implement the mathematical model by a semi-implicit parametric finite element method to study several interesting phenomena, such as ``small'' particle migration on curved substrates and templated solid-state dewetting on a pre-patterned substrate. Based on ample numerical simulations,
we demonstrate that, the migration velocity of a ``small'' solid particle is proportional to the substrate curvature gradient $\hat{\kappa}'$ and inversely proportional to the square root of the area of the particle $\sqrt{A}$, and it decreases when the isotropic Young angle $\theta_i$ increases. In addition, we also observe four periodic categories of dewetting
on a pre-patterned sinusoidal substrate. Our approach can provide a convenient and powerful tool to exploring how to produce well-organized nanoparticles by making use of template-assisted solid-state dewetting.

\end{abstract}

\date{\today}
\maketitle

\section{Introduction}
Solid-state dewetting of thin films has been observed in various thin film/substrate systems by many research groups~\cite{Thompson12,Leroy16,Jiran90,Jiran92,Ye10a,Ye10b,Ye11a,Ye11b,Rabkin14,Kosinova14,Kovalenko17,
Naffouti16,Naffouti17,Pierre09b}, and has attracted increasing attention because of its considerable technological interest.
Especially, in recent years, the solid-state dewetting can be used to provide a simple method for making ordered nanoparticles and quantum dot arrays which have a rich variety of applications, such as used for sensors~\cite{Armelao06, Mizsei93}, optical and magnetic devices~\cite{Armelao06, Rath07}, as catalysis for the growth of carbon and semiconductor nanotube and nanowire \cite{Randolph07, Schmidt09}. Ono et al.~\cite{Ono95} first observed the solid-state dewetting (or agglomeration) in the silicon-on-insulator (SOI) system. Following with the experiment, many experimental studies on dewetting of single crystal films (mostly for SOI~\cite{Nuryadi00, Nuryadi02} and Ni~\cite{Ye10a, Ye10b, Ye11a, Ye11b} films) have been performed and have shown that it could produce well-ordered and controllable patterns. Unlike single crystal films, polycrystalline films usually lead to disordered structures on a flat substrate. While recent experiments have shown that
thin films can evolve into ordered arrays of nanoparticles and well-organized patterns on a pre-patterned substrate, i.e., by making use of the templated solid-state dewetting~\cite{Giermann05,Giermann11,Ye11b,Naffouti16}.
These, and related studies have led to increasing research interests on studying the kinetics of solid-state dewetting of thin films on both flat and curved substrates.

The dewetting of solid thin films deposited on substrates is similar to the dewetting of liquid films~\cite{deGennes85}, and they share some common features, such as the moving contact line~\cite{Qian06,Ren07,Tripathi18}, Rayleigh instability~\cite{Pairam09,Mcgraw10,Kim15}, multi-scale and multi-physics features~\cite{Spencer97,Xu11,Herminghaus08,Khenner18}. However, they have many important major differences. For example, their mass transport processes are totally different, and the solid-state dewetting occurs through surface diffusion instead of fluid dynamics in liquid dewetting; in addition, the surface energy anisotropy plays
an important role in determining equilibrium shapes of particles and the kinetic evolution during the solid-state dewetting, while the isotropic surface energy is usually assumed in liquid dewetting. In the literature, the solid-state dewetting is usually modeled as a surface-tracking problem described by surface diffusion flow, coupled with moving contact lines where the film-vapor-substrate three phases meet with each other~\cite{Srolovitz86b,Wong00,Du10,Dornel06,Jiang12,Wang15,Jiang16}.

Based on different understandings to this problem, there have been lots of theoretical and modeling studies for solid-state dewetting problems in the literature.  Srolovitz and Safran~\cite{Srolovitz86b} first proposed a sharp-interface model to investigate the hole growth under the three assumptions, i.e., isotropic surface energy, small slope profile and cylindrical symmetry. Based on the model, Wong et al.~\cite{Wong00} designed a ``marker particle'' numerical method for solving the two-dimensional fully-nonlinear isotropic sharp-interface model
(i.e., without the small slope assumption), and to investigate the two-dimensional edge retraction of a semi-infinite step film. Dornel et al.~\cite{Dornel06} designed another numerical scheme to study the pinch-off phenomenon of two-dimensional island films with high-aspect-ratios during solid-state dewetting. Jiang et al.~\cite{Jiang12} designed a phase-field model for simulating solid-state dewetting of thin films with isotropic surface energies, and this approach can naturally capture the topological changes that occur during evolution. Although most of the above models are focused on the isotropic surface energy case, recent experiments have clearly demonstrated that the kinetic evolution that occurs during solid-state dewetting is strongly affected by crystalline anisotropy~\cite{Thompson12,Leroy16}. In order to investigate surface energy anisotropy effect, many approaches have been proposed and discussed, such as a discrete model~\cite{Dornel06}, a kinetic Monte Carlo model~\cite{Pierre09b,Dufay11}, a crystalline model~\cite{Carter95,Zucker13} and continuum models based on partial differential equations (PDEs)~\cite{Wang15,Jiang16,Bao17}.

While most of these works are restricted on the flat substrate, dewetting of thin solid films on curved substrates is still not well understood. For simulating template-assisted solid-state dewetting, Giermann and Thompson proposed a simple model~\cite{Giermann11} to semi-quantitatively understand some observed phenomena, but they could not include the contact line/point migration or the surface energy anisotropy into the simple model. Klinger and Rabkin~\cite{Klinger12} developed a discrete algorithm for simulating capillary-driven motion of nanoparticles on curved rigid substrates in two dimensions. In their approach, the self-diffusion along the film/substrate interface (i.e., interface diffusion) and the surface diffusion along the particle surface are included, and the continuity of fluxes and chemical potentials of the interface and surface diffusions at the moving contact point is used to tackle the moving contact line problem. To the best of our knowledge, there are no completed continuum PDE models, which are used for simulating the kinetics of solid particles on curved substrates, available in the literature.


In recent years, a continuum model based on sharp-interface approach was proposed by the authors for simulating solid-state dewetting of thin films on flat substrates~\cite{Wang15,Jiang16,Bao17b} in two dimensions. This continuum model is obtained from the thermodynamic variation to the total interfacial free energy functional and Mullins's method for deriving surface diffusion equation~\cite{Mullins57}. This model describes the interface evolution which occurs through surface diffusion
and contact point migration, and the surface energy anisotropy is easily included into the model, no matter how strong the
anisotropy is, i.e, weakly anisotropic~\cite{Wang15} and strongly anisotropic~\cite{Jiang16}. From mathematics, we can rigorously prove that the sharp-interface model fulfills the area/mass conservation and the total free energy dissipation
properties when following with the kinetics described by the model, and a parametric finite element method was
designed to efficiently solve the mathematical model~\cite{Bao17}. Furthermore, we have extended these approaches to simulating
solid-state dewetting in three dimensions recently~\cite{Bao18,Zhao17thesis}, i.e., moving open surface coupled with moving contact lines. In this paper, we will generalize the modeling techniques and numerical methods to study solid-state dewetting of thin films on non-flat rigid substrates.

In this paper, we assume that the surface diffusion is the only driving force for solid-state dewetting, and that elastic (interface stress, stresses associated with capillarity) effects are negligible, and there are no chemical
reactions or phase transformations occurring during the evolution.
The rest of this paper is organized as follows. In Section II, based on a thermodynamic variational approach, we rigorously derive a mathematical sharp-interface model for simulating solid-state dewetting of thin films on curved rigid substrates. Then, we perform numerical simulations to investigate several specific phenomena about solid-state dewetting of thin films on curved substrates, i.e., the equilibrium shapes of small island films and the pinch-off of large island films in Section III, the ``small'' solid particle migration in Section IV and templated solid-state dewetting in Section V. Finally, we draw some conclusions in Section VI.

\section{Mathematical formulation}

We first discuss the surface evolution kinetics for solid-state dewetting of thin films on rigid, curved substrates in two dimensions (2D). Following the usual non-equilibrium thermodynamic approach, we model the kinetics as driven by the variation of the free energy of the system with respect to matter transport in a sharp-interface framework.

\begin{figure}[!htp]
\centering
\includegraphics[width=.45\textwidth]{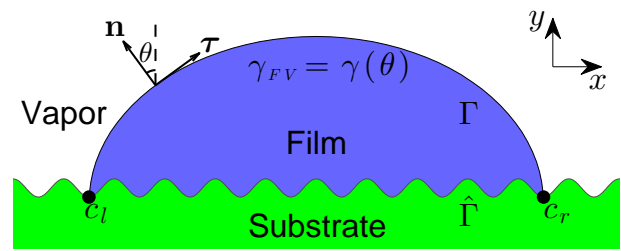}
\caption{A schematic illustration of a solid film (island) in contact with a rigid, curved
substrate in two dimensions, where $c_l$ and $c_r$ represent the left and right contact points,
$\Gamma$ is the film/vapor interface curve, and $\hat{\Gamma}$ is the curved substrate.}
\label{fig:1}
\end{figure}

Most of the relevant variables are described by reference to the example shown in Fig.~\ref{fig:1}.
We denote the film/vapor interface profile as $\Gamma = \mathbf{X}(s) = \big(x(s),y(s)\big), s \in [0,L]$ where $s$ and $L$ represent the arc length and the total length of the interface, respectively.
The unit tangent vector $\bmath{\tau}$ and outer unit normal vector $\mathbf{n}$ of the film/vapor interface curve $\Gamma$ can be expressed as $\bmath{\tau}:=(x_s, y_s)$ and $\mathbf{n}:=(-y_s, x_s)$, respectively.
The angle $\theta$ represents the angle between the local outer unit normal vector and the $y$-axis (or the local tangent vector and the $x$-axis).

The curved rigid substrate profile is denoted as $\hat{\Gamma}:= \hat{\mathbf{X}}(c) = \big(\hat{x}(c), \hat{y}(c)\big)$ with arc length $c\in [0, \hat{L}]$, and $\hat{L}$ represents the total length of the curved substrate. Similarly, $\Tsub$, $\Nsub$ and $\hat{\theta}$ 
represent the unit tangent vector, the (outer) unit normal vector of the curved substrate $\Gammasub$  and the angle between the local unit normal vector and the $y$-axis.

The left and right contact points are located at the intersections of the interface curve $\Gamma$ and the substrate curve $\Gammasub$, i.e., the contact points are at $s = 0$ and $s = L$  on $\Gamma$ and $c = c_l$ and $c = c_r$ on $\Gammasub$. For simplicity, we denote both as $c_l$ and $c_r$ (shown in Fig.~\ref{fig:1}), and  represent the tangent angles to the external surface $\Gamma$ and substrate $\Gammasub$ at the two contact points as
\begin{gather*}
    \theta_{\rm e}^l := \theta(s=0), \quad  \theta_{\rm e}^r := \theta(s=L), \\
    \hat{\theta}^l := \hat{\theta}(c=c_l), \quad \hat{\theta}^r := \hat{\theta}(c=c_r),
\end{gather*}
where $\theta_{\rm e}^l$ and $\theta_{\rm e}^r$ are the left and right extrinsic contact angles~\cite{Xu17}, respectively. Hence, the left and right intrinsic (or true) contact angles are
\begin{equation}\label{eq:def_thetad}
    \theta_{\rm i}^l : = \theta_{\rm e}^l - \hat{\theta}^l,\quad \theta_{\rm i}^r : = \theta_{\rm e}^r - \hat{\theta}^r.
\end{equation}
which satisfy
\begin{equation*}
\cos \theta_{\rm i}^l = \bmath{\tau}(0) \cdot \Tsub(c_l), \qquad \cos \theta_{\rm i}^r = \bmath{\tau}(L) \cdot \Tsub(c_r).
\end{equation*}

Following with the above notations, the total interfacial free energy of the three-phase solid-state dewetting system (including possibly anisotropic surface energies) can be written as~\cite{Wang15,Jiang16,Bao17b}:
\begin{equation}\label{eq:energy}
W=\int_{\Gamma}\gamma(\theta)\;d\Gamma+
\underbrace{\big(\gamma_{\subFS}-\gamma_{\subVS}\big)(c_r-c_l)}_{{\textbf {Substrate\;Energy}}},
\end{equation}
where the first term represents the film-vapor interface energy and the second term represents the substrate interface energy (we have subtracted the energy of the bare substrate).
$\gamma_{\subFV}$, $\gamma_{\subFS}$ and $\gamma_{\subVS}$ are the surface energy densities of the film/vapor, film/substrate and vapor/substrate interfaces, respectively. Here, we assume that $\gamma_{\subFS}$ and $\gamma_{\subVS}$ are two constants, and the film/vapor interface energy density is a function of the interface orientation angle, i.e., $\gamma_{\subFV}:=\gamma(\theta)$.
If $\gamma(\theta)\equiv {\text {constant}}$, the surface energy is isotropic; otherwise, it is anisotropic.
Furthermore, if the surface stiffness $\widetilde\gamma(\theta):=\gamma(\theta)+\gamma^{\prime\prime}(\theta)>0$ for all
$\theta\in[-\pi,\pi]$, the surface energy is weakly anisotropic; otherwise,
if $\widetilde\gamma(\theta)=\gamma(\theta)+\gamma^{\prime\prime}(\theta)<0$ for some orientations $\theta\in[-\pi,\pi]$, the surface energy is strongly anisotropic.

As shown rigorously (and in detail) in Appendix A, the first-order thermodynamic variations of the total free energy $W$ with respect to the film/vapor interface profile $\Gamma$ and the two contact points $c_r$ and $c_l$ are
\begin{eqnarray}
\frac{\delta W}{\delta \Gamma}&=&\Big(\gamma(\theta)+\gamma\,''(\theta)\Big)\kappa,\label{eqn_ch4:var1} \\ [0.6em]
\frac{\delta W}{\delta c_r}&=&\gamma(\theta_{\rm e}^r)\cos\theta_{\rm i}^r-\gamma\,'(\theta_{\rm e}^r)\sin\theta_{\rm i}^r + (\gamma_{\subFS}-\gamma_{\subVS}), \label{eqn_ch4:var2}\\ [0.6em]
\frac{\delta W}{\delta c_l}&=&-\Big[\gamma(\theta_{\rm e}^l)\cos\theta_{\rm i}^l-\gamma\,'(\theta_{\rm e}^l)\sin\theta_{\rm i}^l+
(\gamma_{\subFS}-\gamma_{\subVS})\Big], \quad
\label{eqn_ch4:var3}
\end{eqnarray}
where $\kappa$ is the curvature of the interface curve $\Gamma$.

From the Gibbs-Thomson relation~\cite{Mullins57,Sutton95} (in terms of the curvature, Eq.~\eqref{eqn_ch4:var1}), we can define the chemical potential $\mu$ at any point along the interface curve $\Gamma$.
Variations in the chemical potential along the interface give rise to a material (film) flux along the interface $\vec J$ and the the normal velocity of the film/vapor interface $V_n$~\cite{Wang15,Mullins57}:
\begin{equation}\label{eq:mu}
\mu=\Omega_0\frac{\delta W}{\delta \Gamma}=\Omega_0\Big(\gamma(\theta)+\gamma\,''(\theta)\Big)\kappa = \Omega_0\widetilde{\gamma}(\theta) \kappa,
\end{equation}
\begin{equation}\label{eq:normalvel}
\vec J = -\frac{D_s\nu}{k_B\,T_e}\nabla_s\, \mu,\quad V_n=-\Omega_0 (\nabla_s \cdot \vec J)=\frac{D_s\nu\Omega_0}{k_B\,T_e}\frac{\partial^2 \mu}{\partial s^2},
\end{equation}
where $\nabla_s$ is the surface gradient operator (i.e.,  the derivative with respect to position $s$ along $\Gamma$), $\Omega_0$ is the atomic volume of the film material, $D_s$ is the coefficient of surface diffusion, $\nu$ is the number of diffusing atoms per unit length, and $k_BT_e$ is the thermal energy.
Equations~\eqref{eqn_ch4:var2} and~\eqref{eqn_ch4:var3} are used to construct the equations of motion for the moving contact points in the manner described in~\cite{Wang15,Jiang16},
\begin{eqnarray}
 \frac{d c_l(t)}{d t} &=&-\eta\frac{\delta W}{\delta c_l}, \quad \text{at} \quad c = c_l, \label{eqn_ch4:frelaxationright}\\ [0.5em]
 \frac{d c_r(t)}{d t}&=&-\eta\frac{\delta W}{\delta c_r}, \quad \text{at} \quad c = c_r,
\label{eqn_ch4:frelaxationleft}
\end{eqnarray}
where the constant $\eta \in (0, \infty)$, represents a contact line (or point) mobility.

Next, we nondimensionalize the equations by scaling all lengths by a constant characteristic length scale $R_0$ (e.g., the initial thickness of the thin film layer), energies in terms of the constant, mean surface energy (density) $\gamma_0=\frac{1}{2\pi}\int_{-\pi}^\pi \gamma(\theta)d\theta$, and time by $t_0=R_0^4/(B\gamma_0)$,  where $B:= D_s\nu\Omega_0^2/(k_BT_e)$ is a material constant (the contact line mobility is therefore scaled by $B/R_0^3$).
With these scalings, the above sharp-interface model for the interface evolution  (Eq.~\eqref{eq:normalvel})
becomes
\begin{equation}\label{eq:GE_curve_weak}
    \begin{cases}
        \displaystyle \frac{\partial{\mathbf{X}}}{\partial t}=V_n \mathbf{n} = \frac{\partial^2 \mu}{\partial s^2} \mathbf{n},\\[0.8em]
        \displaystyle \mu=\widetilde{\gamma}(\theta) \kappa = \Big(\gamma(\theta)+\gamma\,''(\theta)\Big)\kappa.
    \end{cases}
\end{equation}
Note that now $\mathbf{X}$, $t$, $V_n$, $s$, $\mu$, $\gamma$, $\kappa$ and $\eta$ are now dimensionless, yet we retain the same notation for brevity.

The dimensionless interface evolution equation~\eqref{eq:GE_curve_weak} is subject to the following dimensionless boundary conditions:
\begin{itemize}
\item[(i)] Contact point condition ({\bf {BC1}})
\begin{equation}\label{eq:BC1_curve_weak}
\mathbf{X}(0, t) = \hat{\mathbf{X}}(c_l), \quad \mathbf{X}(L, t) = \hat{\mathbf{X}}(c_r).
\end{equation}
\end{itemize}
This ensures that the left and right contact points  move along the rigid, curved substrate $\Gammasub$ and simultaneously lie on both the film/vapor $\Gamma$ and substrate $\Gammasub$ interfaces.

\begin{itemize}
\item[(ii)] Relaxed/dissipative contact angle condition ({\bf {BC2}})
\begin{equation}\label{eq:BC2_curve_weak}
\frac{d c_l}{d t} = \eta\, f(\theta_{\rm e}^l, \theta_{\rm i}^l), \qquad \frac{d c_r}{d t} = -\eta\, f(\theta_{\rm e}^r, \theta_{\rm i}^r),
\end{equation}
\end{itemize}
where
\begin{equation*}
    f(\theta_{\rm e}, \theta_{\rm i}) := \gamma(\theta_{\rm e})\cos\theta_{\rm i}-\gamma\,'(\theta_{\rm e})\sin\theta_{\rm i} - \sigma,
\end{equation*}
and $\sigma := (\gamma_{\subVS} - \gamma_{\subFS})/\gamma_0$. The contact angles $\theta_{\rm e}^l, \theta_{\rm e}^r, \theta_{\rm i}^l, \theta_{\rm i}^r$  are related as per Eq.~\eqref{eq:def_thetad} and hence are intrinsically related to the substrate shape.

\begin{itemize}
\item[(iii)] Zero-mass flux condition ({\bf {BC3}})
\begin{equation}\label{eq:BC3_curve_weak}
\frac{\partial \mu}{\partial s}(0, t)=0, \qquad \frac{\partial \mu}{\partial s}(L, t)=0,
\end{equation}
\end{itemize}
This condition implies that the total mass of the film is conserved (see Appendix B).

If the film evolves to a stationary state, the contact angles evolution equation~\eqref{eq:BC2_curve_weak} ensures that the equilibrium contact angle is achieved by
$\gamma(\theta_{\rm e})\cos\theta_{\rm i}-\gamma\,'(\theta_{\rm e})\sin\theta_{\rm i} = \sigma$ .
This is the classical Young equation generalized for the curved substrate case.
If the  surface energy is isotropic (i.e., $\gamma(\theta) \equiv 1$, and $\gamma\,'(\theta) \equiv 0$), the generalized Young equation reduces to the classical isotropic Young equation~\cite{Young1805}, i.e., $\cos\theta_{\rm i}=\sigma$.
On the other hand, when the substrate is flat ($\hat{\theta}\equiv 0$), the generalized Young equation reduces to the classical anisotropic Young equation~\cite{Wang15, Jiang16} (in this case $\theta_{\rm e} = \theta_{\rm i}$). However, when the substrate is curved, we cannot, in general, explicitly determine the static intrinsic angles for arbitrary anisotropy.

We demonstrate, in Appendix B, that the general (anisotropic) evolution equation \eqref{eq:GE_curve_weak} together with boundary conditions \eqref{eq:BC1_curve_weak}-\eqref{eq:BC3_curve_weak} ensures that the total film mass (area) is conserved and the total free energy of the system decreases monotonically during film morphology evolution. From a mathematical point of view, we note that the governing equations are well-posed when the surface energy is isotropic or weakly anisotropic. On the other hand, when the surface energy is strongly anisotropic, the equations will become of the anti-diffusion type (e.g., likewise, a second-order diffusion term with a negative ``diffusion'' coefficient) and are ill-posed. We handle this ill-posedness by regularizing the equations by adding high-order terms (e.g., see~\cite{Jiang16}).

\section{Island evolution on curved substrates}

We employ a parametric finite-element method to numerically solve the above mathematical model for the evolution of islands on curved substrates. The numerical algorithm is described in Appendix C and was previously applied to solid-state dewetting problems on flat substrates in~\cite{Bao17}.
Our numerical examples all use an anisotropic film/vapor surface energy (density) of the following form
\begin{equation}
    \gamma(\theta) = 1+\beta\cos(m\theta),
\end{equation}
where the parameter $\beta$ controls the degree of the anisotropy and $m$ describes the order of the rotational symmetry.
For $\beta=0$, the surface energy is isotropic.
For $0<\beta<\frac{1}{m^2-1}$,
it is weakly anisotropic.
And, for $\beta>\frac{1}{m^2-1}$, it is strongly anisotropic.
We focus here on the case of large contact point mobility ($\eta=100$).
A more detailed discussion of the influence of the parameter $\eta$ and contact-line drag on the kinetic evolution process (and even stationary morphologies) can be found in~\cite{Wang15}.

\subsection{Small island equilibrium }

\begin{figure}
\centering
\includegraphics[width=.48\textwidth]{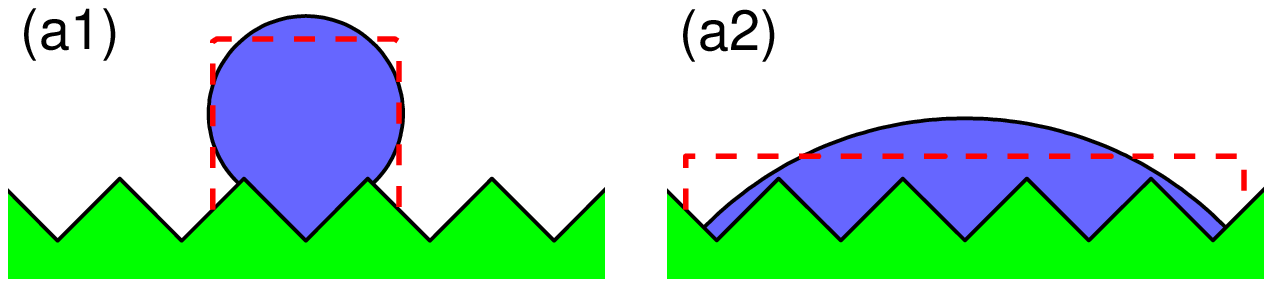}
\includegraphics[width=.48\textwidth]{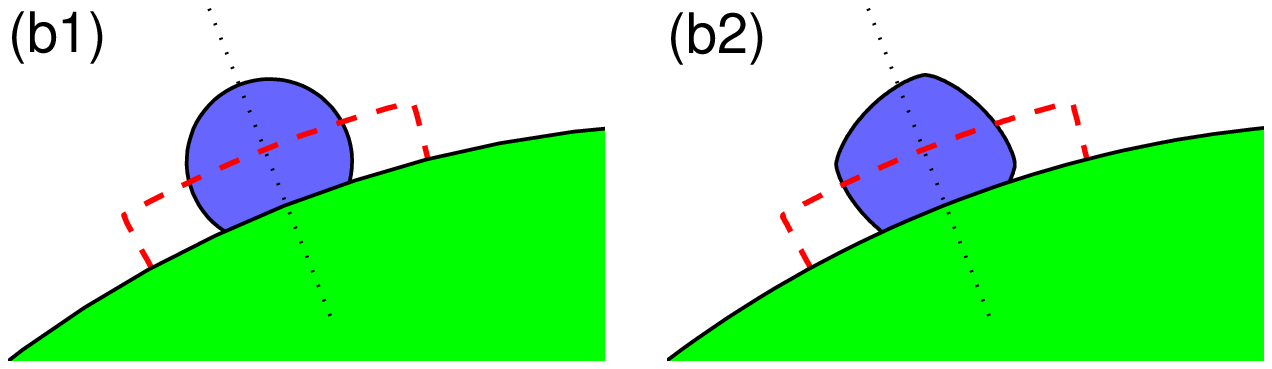}
\caption{(a1-a2) show two equilibrium isotropic islands with material constant $\sigma = 0$ (intrinsic contact angles are both $\pi/2$) on a sawtooth substrate starting from two different initial island shapes (indicated by the red dashed lines); (b1-b2) shows two equilibrium shapes of island films with material constant $\sigma = -0.5$ on a circular substrate with radius $R = 20$, where (b1) is the isotropic case with static intrinsic contact angle $2\pi/3$, (b2) is the weakly anisotropic case (where $m = 4, \beta = 0.06$) with static intrinsic contact angles $2.025$ (left) and $2.319$ (right).}
\label{fig:3-1}
\end{figure}

Isotropic islands on flat substrates evolve to the same stationary state determined by the equilibrium contact angle, independent of the initial island shape.
However, this is not necessarily the case when the substrate is not flat, as illustrated in Fig.~\ref{fig:3-1}~(a1-a2) for the case of a sawtooth-profile substrate.
Here, the stationary island shapes (evolving from different initial island shapes) have very different macroscopic aspect ratios and cover vastly different substrate lengths (areas).
This suggests the possibility of manipulating island shape through control of substrate morphology and/or  initial island profile.

Fig.~\ref{fig:3-1} (b1-b2) shows two stationary island shapes for islands on a circular substrate with exactly the same values of the  material parameter $\sigma$. In the first case, the island surface energy is isotropic, while in the second the surface energy is weakly anisotropic. Initially, the two islands have the same shapes and locations. As can be clearly seen from the figure, the isotropic island evolves to a symmetric circular shape with static intrinsic contact angle $2\pi/3$; while the anisotropic island evolves to an asymmetric island shape (the shape itself is determined by the surface energy anisotropy) and has two different left and right static intrinsic contact angles. These numerical results indicate that the surface energy anisotropy  can lead to multiple static intrinsic contact angles on curved substrates. The presence of different (left and right) contact angles on the same island was observed earlier for strongly anisotropic islands on a flat substrate but not for weakly anisotropic islands~\cite{Bao17b}. This feature of weakly anisotropic islands is associated with the fact that here, the substrate is curved.

\subsection{Large island pinch-off }

\begin{figure}
\centering
\includegraphics[width=.49\textwidth]{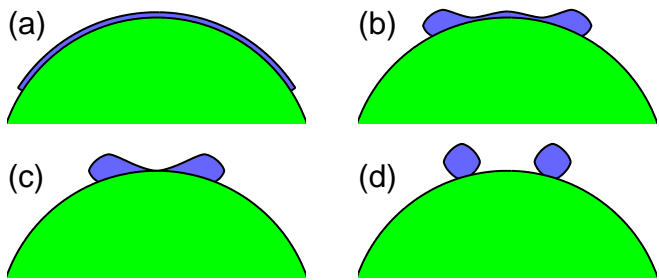}
\caption{Morphology evolution of a large island film (aspect ratio $L=60$) with weakly anisotropic surface energy on a circular substrate of radius $R = 30$, where  $m = 4, \beta = 0.06, \sigma = -\sqrt{3}/2$.}
\label{fig:3-2}
\end{figure}

When the aspect ratio of an island film is larger than a critical value, the island will pinch off and break up into two or more islands. In analogy  to  pinch-off  on flat substrates~\cite{Dornel06,Wang15}, we perform numerical simulations of large islands on circular curved substrates.
Fig.~\ref{fig:3-2} shows several configurations during the  evolution  of a large-aspect-ratio island  
on a circular substrate of radius $R=30$.
As shown in Fig.~\ref{fig:3-2}, surface diffusion  very quickly leads to the formation of ridges at the island edges followed by valleys; then as time evolves, the two valleys merge near the island center; eventually, the valley at the center of the islands deepens until it touches the substrate, leading to a pinch-off event that separates the initial island into a pair of islands. This  evolution  is very similar to that on flat substrates~\cite{Wang15}.

\begin{figure}[!htp]
\centering
\includegraphics[width=.4\textwidth]{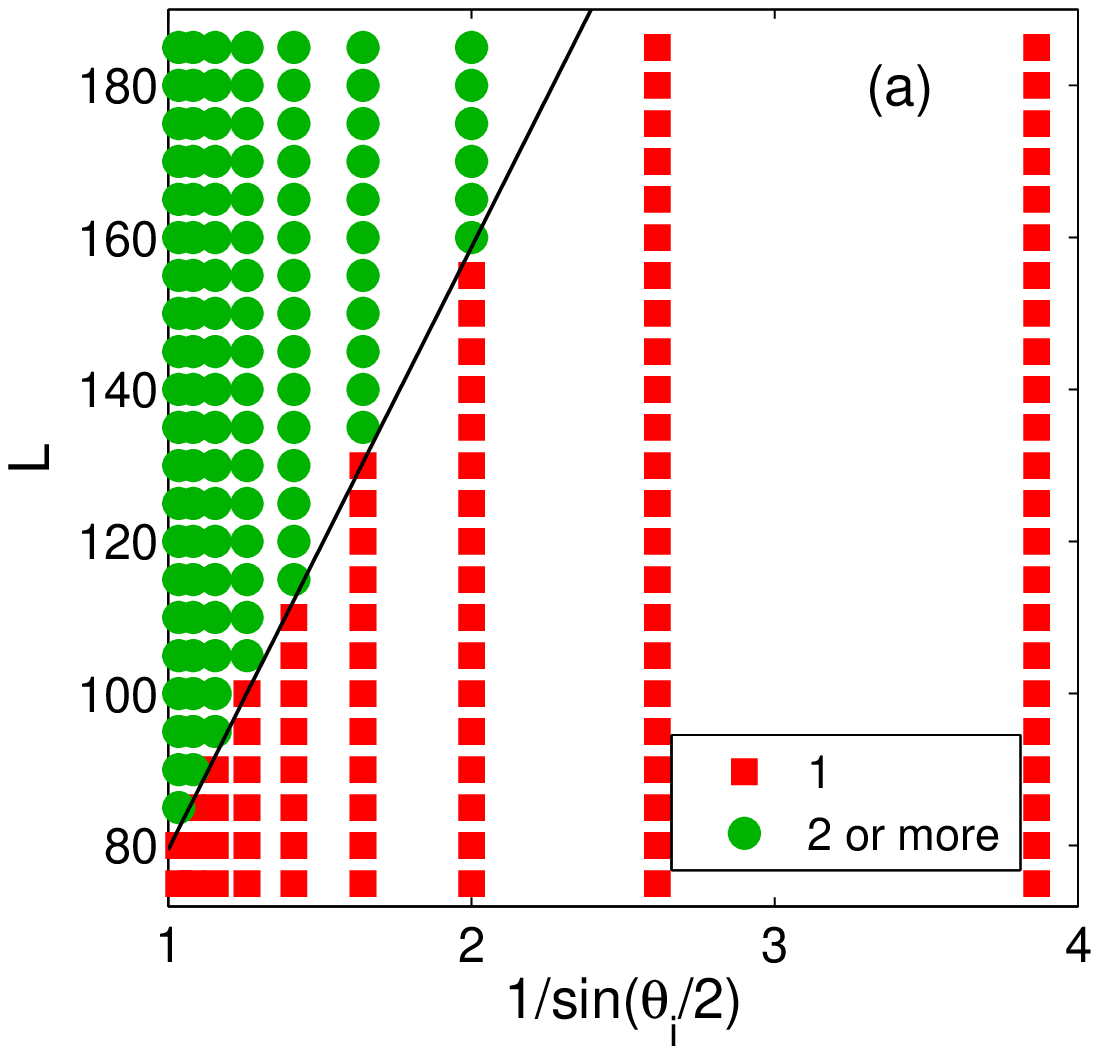}
\includegraphics[width=.4\textwidth]{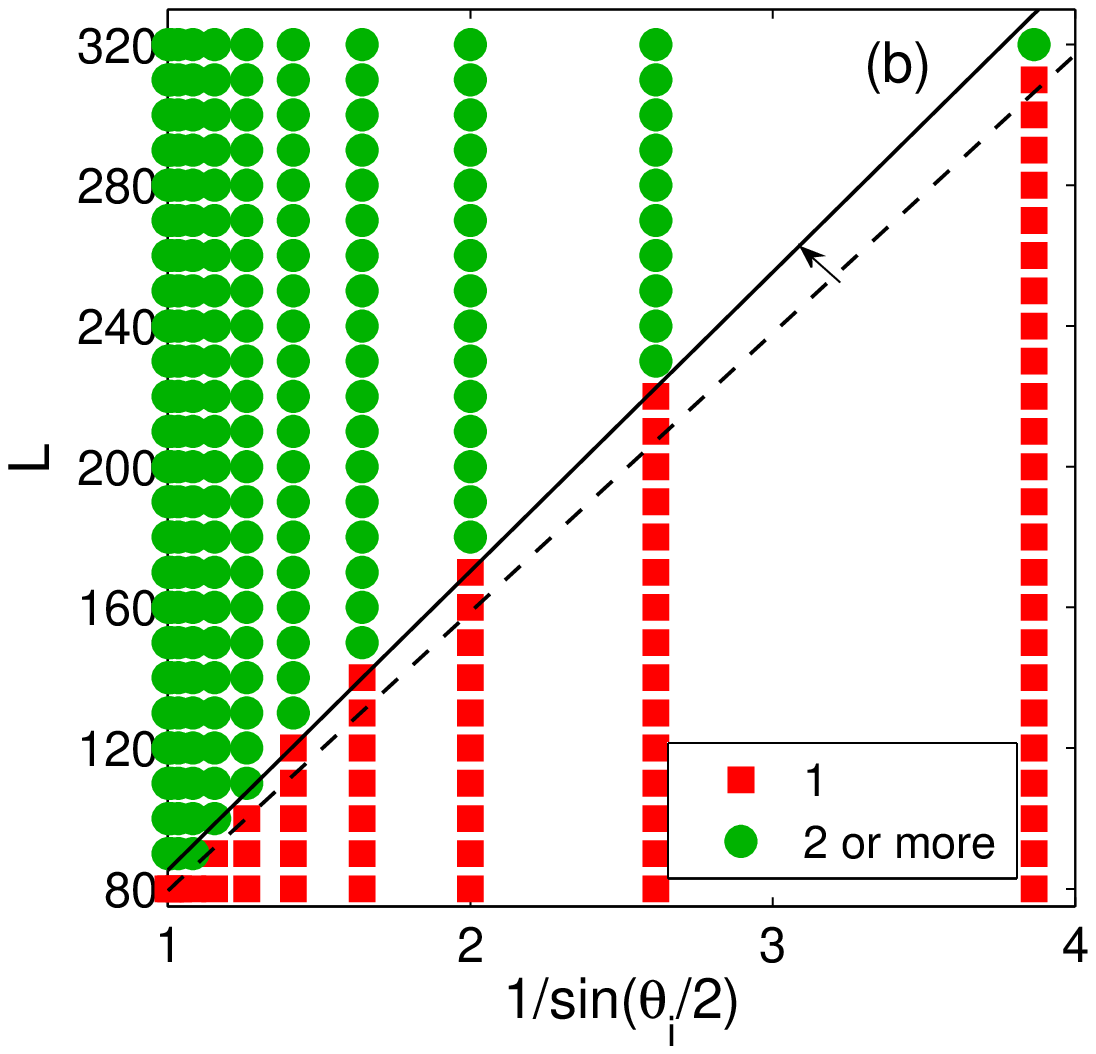}
\caption{The number of islands formed from the retraction of a high-aspect-ratio island (with isotropic Young angle $\theta_i$; $\sigma=\cos\theta_i$) as a function of initial length $L$ on circular substrates of radii (a) $R = 30$ and (b) $R = 60$. The solid black lines separating the one and two island domains correspond to (a) $L_c = 79.2/\sin(\theta_i/2) + 0.2$, (b) $L_c = 85.0/\sin(\theta_i/2) + 0.3$. The black dashed line in (b) is the solid black line in (a).}
\label{fig:3-3}
\end{figure}

We now  investigate how the substrate curvature affects the critical pinch-off length $L_c$ of island films (above which  pinch-off occurs). Fig.~\ref{fig:3-3} shows the number of small islands formed during  solid-state dewetting on circular substrates of  radii $R = 30$ and $60$ for isotropic surface energy and Young angles $\theta_i\in [0, \pi]$. This shows that the boundary line separating domains of different number of pinched off islands is well-fitted by straight lines: $L_c=79.2/\sin(\theta_i/2)+0.2$ for $R=30$ and $L_c = 85.0/\sin(\theta_i/2)+0.3$ for $R = 60$, respectively.  We  performed similar calculations for substrates of several curvatures and intrinsic contact angle $\theta_i$.  The resultant critical pinch-off lengths for different $R$ and $\theta_i$ are shown in Table~\ref{tab_ch4:length} (the flat substrate result $R\to\infty$ is obtained from the fitting formula of Dornel~\cite{Dornel06}).
This table shows that the critical pinch-off length increases with decreasing isotropic Young angle $\theta_i$  and increasing substrate radius $R$.
We fit these numerical results for the critical pinch-off film length $L_c$ (as a function of isotropic Young angle $\theta_i$ and substrate radius $R$) to the functional form
\begin{equation}
\label{eqn:curve}
L_c = \frac{a(R)}{\sin(\theta_i/2)} + b(R),
\end{equation}
where the functions $a(R)$ and $b(R)$ are well approximated by $a(R) \approx -320.2/R + 89.9$ and $b(R)\approx 0.0$ for $R\ge 20$.

\begin{table}[!htp]
\renewcommand\arraystretch{1.4}
  \centering
  \begin{tabular}{l||c|c|c|c|c|c}
    \hline
    \hline
     & $R = 20$ & $R = 30$ & $R = 40$ & $R = 50$ & $R = 60$ & $R \to \infty$ \\
      \hline
    $\theta_i = \pi$ &  73.5 & 77.5 & 79.5 & 80.5 & 81.5 & 87.9\\
    \hline
    $\theta_i = \frac{11}{12}\pi$ &  74.5 & 78.5 & 80.5 & 81.5 & 82.5 & 88.8\\
    \hline
    $\theta_i = \frac{10}{12}\pi$ &  76.5 & 81.5 & 83.5 & 84.5 & 84.5 & 91.3\\
    \hline
    $\theta_i = \frac{9}{12}\pi$ &  80.5 & 85.5 & 87.5 & 88.5 & 89.5 & 95.9\\
    \hline
    $\theta_i = \frac{8}{12}\pi$ &  86.5 & 91.5 & 94.5 & 95.5 & 96.5 & 102.9  \\
    \hline
    $\theta_i = \frac{7}{12}\pi$ &  94.5 & 100.5 & 103.5 & 105.5 & 106.5 & 113.1 \\
    \hline
    $\theta_i = \frac{6}{12}\pi$ &  105.5 & 113.5 & 119.5 & 119.5 & 121.5 & 128.0 \\
    \hline
    $\theta_i = \frac{5}{12}\pi$ &  120.5 & 131.5 & 137.5 & 140.5 & 142.5 & 150.0 \\
    \hline
    $\theta_i = \frac{4}{12}\pi$ &  -- & 157.5 & 166.5 & 170.5 & 172.5 & 184.5 \\
    \hline
    $\theta_i = \frac{3}{12}\pi$ &  -- & -- & 210.5 & 219.5 & 224.5 & 243.8 \\
    \hline
    $\theta_i = \frac{2}{12}\pi$ &  -- & -- & -- & 306.5 & 319.5 & 364.6 \\
    \hline
    \hline
  \end{tabular}
  \caption{Critical  island film length $L_c$ for island break-up as a function of  isotropic Young angles $\theta_i$
  (i.e., the material constant $\sigma=\cos\theta_i$) and substrate radius $R$ for the isotropic surface energy case. The symbol ``-'' implies that no pinch-off occurred (i.e., $L_c>2\pi R$). The $R\to\infty$ (flat substrate) data is consistent with earlier results~\cite{Dornel06}.}
  \label{tab_ch4:length}
\end{table}

\section{Migration of ``small'' islands}

In this section, we will examine the evolution of small islands on substrates with non-constant surface curvature.
As discussed above (see Section III.1), the equilibrium shape of small islands on substrates with constant surface curvature for both the cases of isotropic and anisotropic surface energies can be determined.
Interestingly, when the substrate curvature is not constant, island migration is possible.
Using a simple model, Ahn and Wynblatt showed that a solid particle will migrate from convex to concave substrate sites~\cite{Ahn80}. Klinger and Rabkin, using a different algorithm, examined the motion of (for example) a particle on a substrate with a sinusoidal profile~\cite{Klinger12}.
Here, we apply the proposed mathematical model to investigate the motion of a ``small''  solid particle on an arbitrarily curved substrate for the case of isotropic surface energy.
As we discuss below, ``small'' implies that the product of the island size (i.e., the area of particle in 2D) and the substrate curvature gradient is small compared with one. This implies that the relaxation time of the island shape is small compared with the time necessary for the island to translate by an island radius.

\begin{figure}[htp!]
\centering
  \includegraphics[width = .48\textwidth]{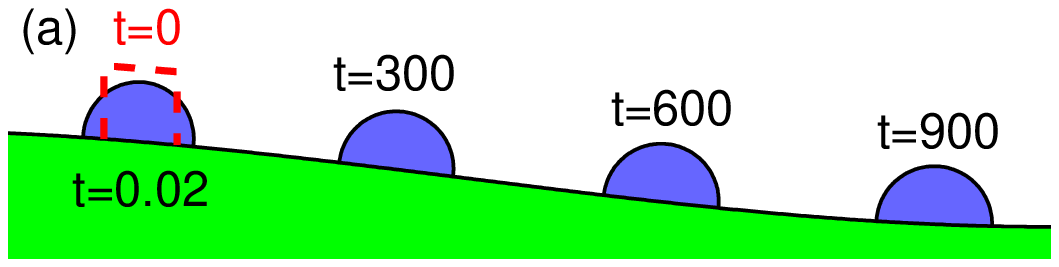}
  \includegraphics[width = .40\textwidth]{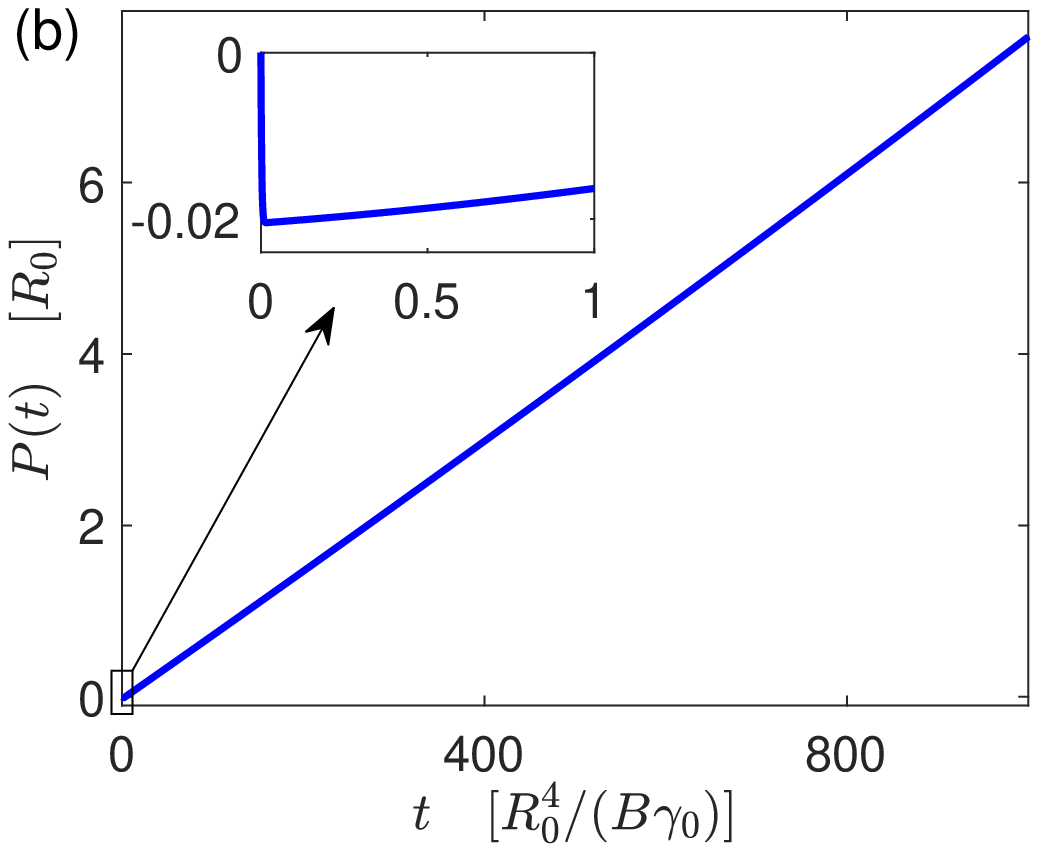}
  \caption{(a) Simulation results for a ``small'' solid particle migration on a curved rigid substrate with a constant curvature gradient $\hat{\kappa}'(c)\equiv -0.01$ at different times $t=0, 0.02, 300, 600, 900$, respectively, where the isotropic Young angle is chosen as $\theta_i = \pi/2$, and the red dashed line represents the initial shape and location of the solid particle (its area $A=0.4$); (b) simulation results for the position of the particle $P(t)$ as a function of time.}
  \label{fig:4-1}
\end{figure}

Here we focus on the leading-order term in the expansion of the total free energy variation that gives rise to particle migration~\cite{Jiang18}; that is, we focus on the effect of a substrate curvature gradient (i.e., $\hat{\kappa}'(c) \equiv {\text{Const.}}$) on the evolution of the particle on the substrate (we assume that $\hat{\kappa}$ is positive for a convex substrate curve). Fig.~\ref{fig:4-1}(a) shows several images during the kinetic evolution of a small, initially square, solid island evolving on a substrate with $\hat{\kappa}' = -0.01$; the evolution was determined by numerical solutions of the proposed sharp-interface model. The position of the particle versus time, $P(t):=(c_l(t)+c_r(t))/2$, is shown in Fig.~\ref{fig:4-1}(b). As is clearly shown that, the island rapidly evolves from its initial square shape (red dashed line) into a nearly perfect circular arc (blue shape, at about $t=0.02$) in an instant time. After the island achieves its near equilibrium shape, it slowly migrates down along the substrate (translates to the right in Fig.~\ref{fig:4-1}). During the migration, the island keeps with its near equilibrium shape. Here, we refer to the time period associated with the island morphology relaxation to its near equilibrium shape as the relaxation time $\tau_R$, and it may be estimated from the inset of Fig.~\ref{fig:4-1}(b), and we estimate this time $\tau_R$ to be around $10^{-2}$.

Since the capillarity-driven evolution is dictated by Eq.~\eqref{eq:GE_curve_weak} (fourth-order in space, first-order in time), the characteristic island shape evolution time $\sim R_0^4$, where $R_0\sim\sqrt{A}$ is the nominal island radius. We demonstrate below that the island translation velocity is proportional to the substrate curvature gradient and inversely proportional to the nominal island radius $R_0$. This implies that the shape evolution rate is much faster than the particle translation rate, when $|A\hat{\kappa}'|\ll1$.  This is the case for the results shown in Fig.~\ref{fig:4-1} ($|A\hat{\kappa}'|=0.004$). Since the relaxation time is small compared with the time required for the island to move an island radius, it is reasonable to assume that the particle shape is always in equilibrium at the local substrate site~\cite{Jiang18}.

\begin{figure}[htp!]
\centering
  \includegraphics[width = .45\textwidth]{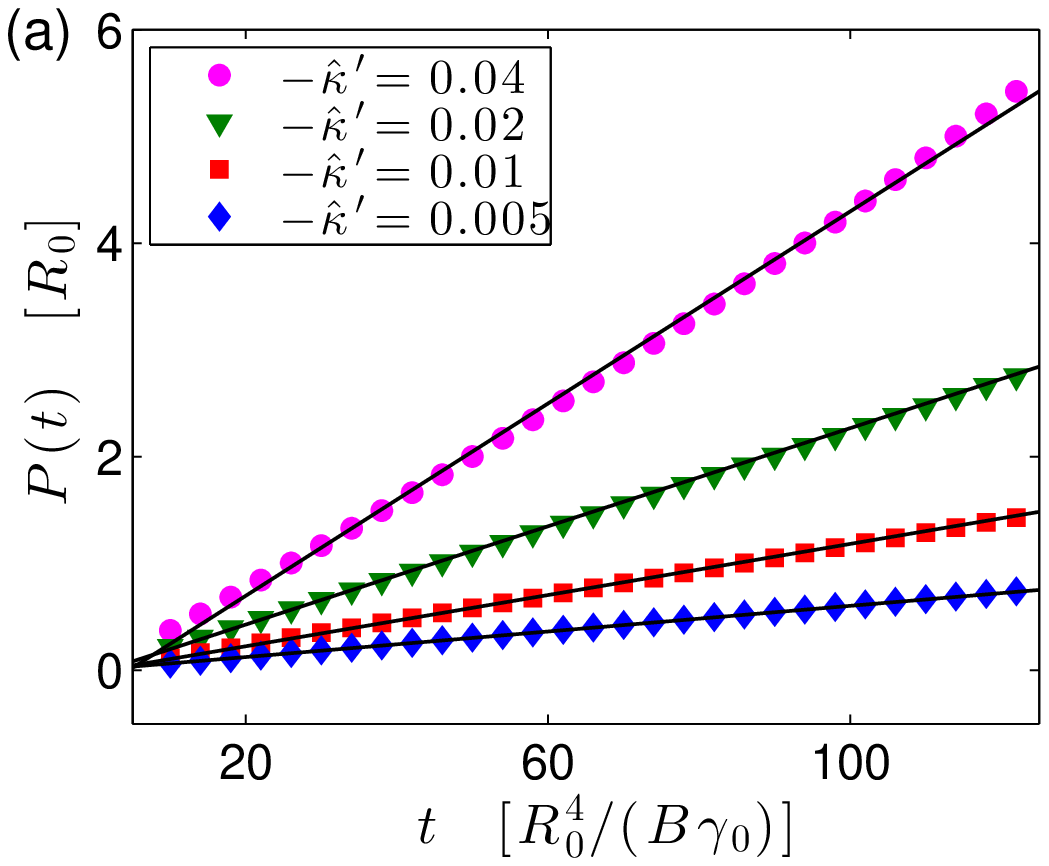}
  \includegraphics[width = .45\textwidth]{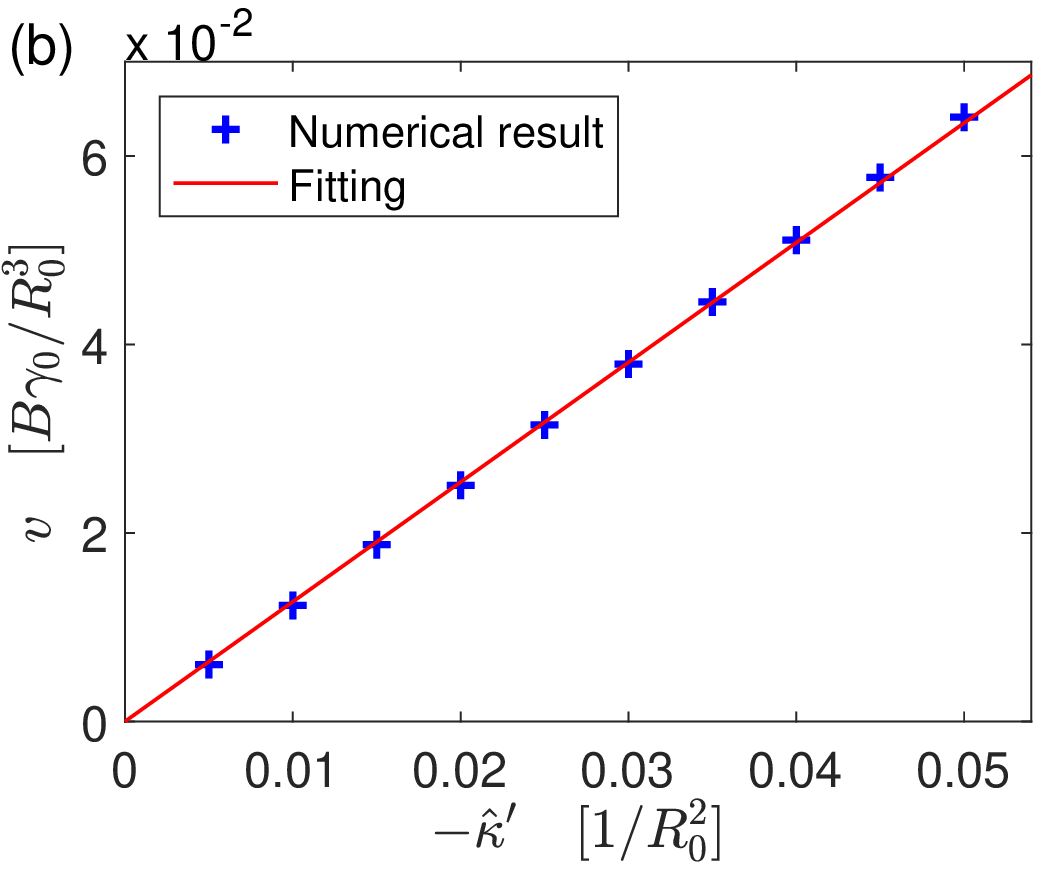}
  \caption{(a) Plot of the position of the ``small'' solid island on the substrate as a function of time for different values of the substrate curvature gradient $\hat{\kappa}'$, where the black solid lines are  least square linear fits to the numerical simulation data (points). (b) Plot of the island velocity as a function of the curvature gradient $\hat{\kappa}'$. These data are well fit by the expression $v = -1.27\,\hat{\kappa}'$ (in red solid line).  In all of these numerical simulations, we fix the island area to be $A = 1$ and the isotropic Young angle to be $\theta_i = \pi/3$.}
  \label{fig:4-2}
\end{figure}

We now examine how the island velocity $v$ varies with substrate curvature gradient $\hat{\kappa}'$, the island area $A$  and the isotropic Young angle $\theta_i$ (i.e., the material constant is chosen as $\sigma=\cos\theta_i$). Numerical simulations were performed for several values of the substrate curvature gradient at fixed island area $A=1$ and Young angle $\theta_i = \pi/3$, and Fig.~\ref{fig:4-2}(a) shows the particle position $P(t)$ versus time. These data are well fit by straight lines, where the slope is a function of substrate curvature gradient $\hat{\kappa}'$; i.e., the particle velocity is nearly constant after a very short time transient (shown in Fig~\ref{fig:4-1}(b)). Least square linear fits to these data yield island velocity versus substrate curvature gradient $\hat{\kappa}'$ as shown in Fig~\ref{fig:4-2}(b). This plot demonstrates that ``small'' island velocity is proportional to the substrate curvature gradient $\hat{\kappa}'$.

\begin{figure}[htp!]
\centering
  \includegraphics[width = .45\textwidth]{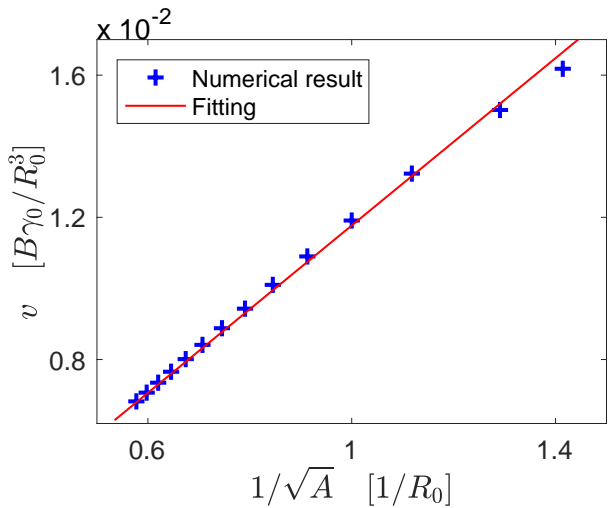}
  \caption{Plot of the island velocity as a function of $1/\sqrt{A}$. These data are well fit by the linear relation $v = 0.01/\sqrt{A}$ (blue solid line). In all of these numerical simulations, we fix the  substrate curvature gradient to be $\hat{\kappa}'= -0.01$ and the isotropic Young angle to be $\theta_i = \pi/3$.}
  \label{fig:4-3}
\end{figure}

\begin{figure}[htp!]
\centering
  \includegraphics[width = .45\textwidth]{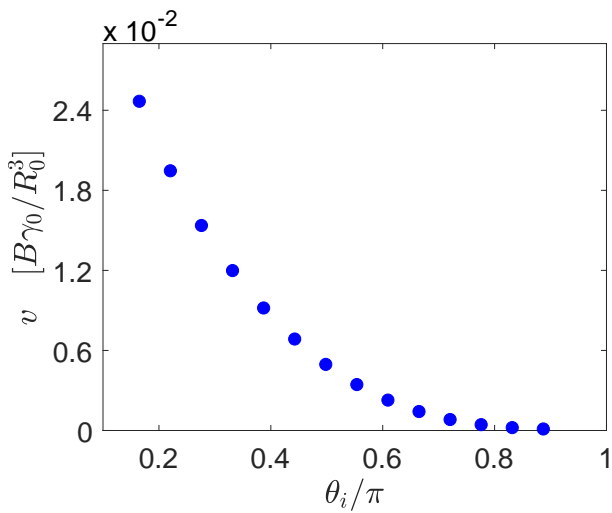}
  \caption{Plot of the island velocity as a function of the isotropic Young angle $\theta_i$.
  In all of these  numerical simulations, we set the substrate curvature gradient to be $\hat{\kappa}'= -0.01$ and the initial island area to $A = 1$.}
  \label{fig:4-4}
\end{figure}

We also examined the relation between the island velocity and the initial island area  $A$ and Young angle $\theta_i$.
The numerical simulation results for the effect of island size are shown in Fig.~\ref{fig:4-3} for a constant substrate curvature gradient  $\hat{\kappa}'= -0.01$ and an isotropic Young angle $\theta_i = \pi/3$.
These data demonstrate that  the ``small'' island velocity  is inversely proportional to the island radius (or more precisely the square root of the island area $\sqrt{A}$), although there are small deviations from this relation for very small islands. The numerical simulation results for the effect of isotropic Young angle $\theta_i$ is shown in Fig.~\ref{fig:4-4}  for fixed curvature gradient $\hat{\kappa}'= -0.01$ and fixed island size $A=1$.
The island velocity increases with decreasing Young angle $\theta_i$ and decreases to zero as $\theta_i \to \pi$. The latter observation is consistent with the fact that a completely dewetting island ($\theta_i = \pi$) will not cover the substrate and hence its free energy is independent of the location where it stands on the curved substrate.

\begin{figure}[hbp!]
\centering
  \includegraphics[width = .48\textwidth]{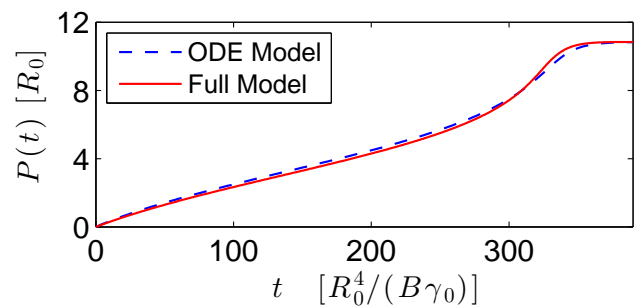}
  \caption{Comparison between solving the full model and the ODE model (i.e., Eq.~\eqref{eq:ode}) for obtaining
  the position of a ``small'' particle at different times during the migration time on a sinusoidal substrate $\hat{y} = 4\sin(\hat{x}/4)$, where the red line represents the numerical result by solving the full model, i.e., Eq.\eqref{eq:GE_curve_weak} together with the boundary conditions \eqref{eq:BC1_curve_weak}-\eqref{eq:BC3_curve_weak}, and the blue dashed line represents the numerical results by solving the ODE model, i.e., Eq.~\eqref{eq:ode}, with $C(\theta_i) = 1.2$. The other parameters are chosen as $A = 1, \theta_i = \pi/3$.}
  \label{fig:4-5}
\end{figure}

Based upon the  numerical results presented here, we  conclude that the migration velocity of ``small'' solid islands
on curved substrates  are well described by the following relation:
\begin{equation}\label{eq:ode}
    v(t):= \frac{{\rm d} P(t)}{\rm d t} = -B \gamma_0 C(\theta_i)\frac{\hat{\kappa}'(P)}{\sqrt{A}},
\end{equation}
where $B:= D_s\nu\Omega_0^2/(k_BT_e)$ is a material constant, $\gamma_0$ is the isotropic particle surface energy density, $C(\theta_i)$ is a function of the isotropic Young angle $\theta_i$ that decreases with increasing $\theta_i$, and $\hat{\kappa}'(P)$ is the local substrate curvature gradient at the arc-length point $P$ on the curved substrate, where $P\in[0, \hat{L}]$ is the arc length along the curved substrate.
In a forthcoming paper~\cite{Jiang18}, based upon the Onsager's variational principle, we can obtain an analytical expression for the function $C(\theta_i)$, which is consistent with the above numerical results.

\begin{figure*}[htp]
\centering
\includegraphics[width=.92\textwidth]{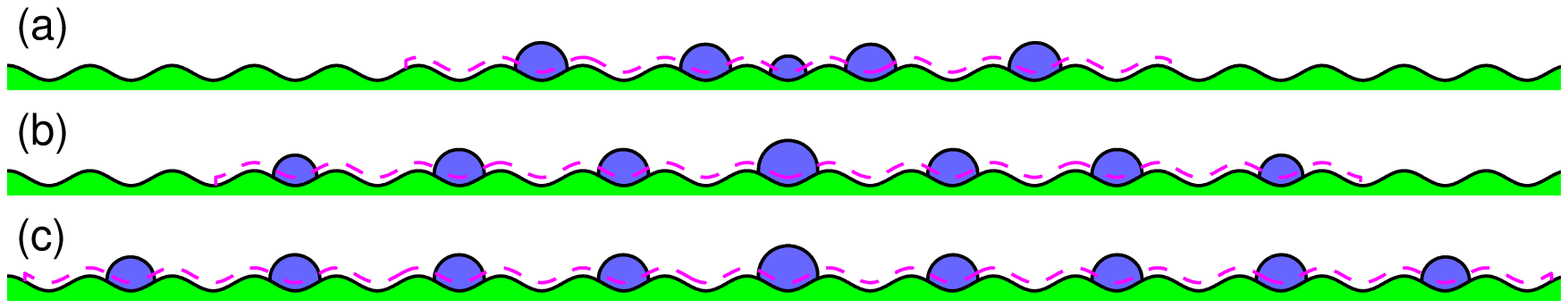}
\caption{Solid-state dewetting of thin film with different initial lengths on a pre-patterned sinusoidal substrate, where the initial length of thin film is chosen as $100$, $150$ and $200$, respectively, and the length scale $R_0$ is chosen as the initial thickness of the thin film.  The magenta dashed line is the initial shape of the thin film, and the shaded blue region is the final equilibrium pattern.}
\label{fig:5-1}
\end{figure*}

While the above numerical results focussed on substrate of fixed curvature gradients, we can characterize an arbitrary substrate profile by a position-dependent substrate curvature gradient $\hat{\kappa}'(P)$.
Hence, since we can determine the velocity of a  ``small'' solid particle at any point along the substrate and by numerically solving the ordinary differential equation in \eqref{eq:ode}, we can predict the trajectory of a ``small'' solid particle on a substrate surface of arbitrary shape. To validate this approach, we numerically simulate the migration of ``small'' solid particles ($A = 1, \theta_i = \pi/3$) on a sinusoidal substrate $\hat{y} = 4\sin(\hat{x}/4)$. The results are shown in Fig.~\ref{fig:4-5}, where the red line represents the results of the numerical simulation via the full model, i.e., Eq.\eqref{eq:GE_curve_weak} together with the boundary conditions \eqref{eq:BC1_curve_weak}-\eqref{eq:BC3_curve_weak}, while the blue dashed line represents the solution of the ordinary differential equation in Eq.~\eqref{eq:ode} for $C(\pi/3) = 1.2$  (see Fig.~\ref{fig:4-4}). These results show the excellent agreement between our ordinary differential equation model Eq.~\eqref{eq:ode} and the numerical solution to the full model.

\section{Templated solid-state dewetting}

In this section, we will apply the sharp-interface model to simulate templated solid-state dewetting on a pre-patterned substrate. The recent experiments have demonstrated that templated solid-state dewetting can be used to controllably produce complex and well-ordered patterns~\cite{Thompson12,Ye11b,Giermann05,Giermann11}. For example, Giermann and Thompson used topographically patterned substrate to modulate the curvature of thin gold films, creating the instabilities which is driven by the solid-state dewetting and results in well-ordered patterns and almost-uniform size of particles, and furthermore,
they observed four general type of island morphologies on this inverted pyramid topography~\cite{Giermann05}.
In a companion paper~\cite{Giermann11}, they proposed two simple models to semi-quantitatively understand the observed phenomena. In this section, we choose the pre-patterned substrate as the sinusoidal curve, which is expressed as $\hat{y}=H\sin(\omega \hat{x})$ with amplitude $H$ and frequency $\omega$, and apply the proposed sharp-interface model to investigate the relation between different type of periodic patterns and the substrate parameters (i.e., $H$ and $\omega$).

Fig.~\ref{fig:5-1} depicts how the finite (initial) length of thin film takes influence on the equilibrium pattern. As shown in the figure, the finite length of thin film will result in non-periodic patterns due to the edge effect, but when the initial length is chosen to be longer and longer, its equilibrium shape will become closer and closer to a periodic pattern. Note that during numerical simulations, when a pinch-off event happens, a new contact point is generated; then after the pinch-off event, we compute each part of the pinch-off curve separately.

\begin{figure*}[htp!]
\centering
\includegraphics[width=.92\textwidth]{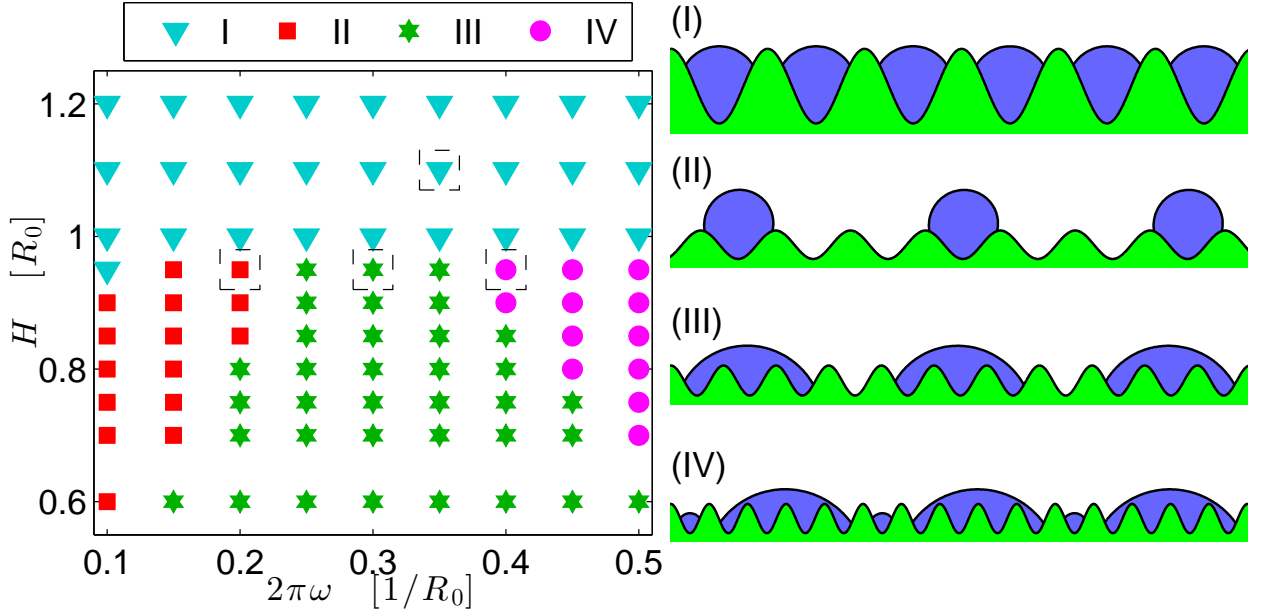}
\caption{Phase diagram of the four observed periodic categories of solid-state dewetting on a pre-patterned sinusoidal substrate, which are: (I) one particle per pit with no empty intermediate pits, (II) one particle occupies one pit with empty intermediate pits, (III) one particle occupies multiple pits with empty intermediate pits, (IV) different sizes of particles. In all above numerical simulations, the isotropic Young angle $\theta_i = 2\pi/3$ and the initial length of thin film is chosen to be long enough.}
\label{fig:5-2}
\end{figure*}

In the following, we performed numerical simulations to investigate the relation. In order to consider the ``periodic'' equilibrium pattern, we choose the initial length of thin films to be long enough. This is the common case, because thin films often have very large aspect ratios. As shown in Fig.~\ref{fig:5-2}, we divide the observed periodic equilibrium patterns into the following four categories of dewetting on a sinusoidal substrate: (I) one particle per pit with no empty intermediate pits; (II) one particle occupies one pit with empty intermediate pits; (III) one particle occupies multiple pits with empty intermediate pits; (IV) different sizes of particles.

The phase diagram of the four periodic categories of dewetting is also depicted in Fig.~\ref{fig:5-2}. As shown in the phase diagram, when the amplitude $H>R_0$ (where $R_0$ is the initial thickness of thin film, and is chosen as the length scale), the equilibrium pattern will fall into the category (I). This can be explained as because the thin film tends to flatten in order to minimize the total interfacial free energy, and if the amplitude of the sinusoidal substrate is too large, it will touch the substrate before flattening and result in one particle in each pit. A simple model~\cite{Giermann11} was proposed to predict the critical amplitude of the substrate, i.e.,
the condition in which the area of thin film is equal to the area of one pit. Here, for a sinusoidal substrate, by some simple calculations, the initial area of the film in one pit is $2\pi R_0/\omega$, and the area of one pit is $2\pi H/\omega$. If they are equal, the critical amplitude is $R_0$, which is excellently consistent with our numerical results.

On the other hand, as shown in Fig.~\ref{fig:5-2}, when $H<R_0$, the equilibrium pattern will fall into three
possible categories: (II)-(IV). In these categories, (II) and (III) are both uniform size of particles, and the intermediate space between these particles can be well-controlled by adjusting the parameters $H$ and $\omega$. When the amplitude $H$ is fixed and the frequency $\omega$ increases to be larger than a critical value, the final pattern will fall into the category (IV), i.e., non-uniform size of particles will appear. Numerical simulations indicate that this critical frequency increases
as the amplitude $H$ decreases, and when ${H}/{R_0}$ goes to zero, the critical frequency will go to infinity. Furthermore, in this case (i.e., $H/R_0\ll 1$), our numerical simulations have demonstrated that the periodicity of the final equilibrium pattern is very close to the one predicted by Wong {\it et al.} in their ``mass-shedding model'' for a thin film on a planar substrate~\cite{Wong00}.

\section{Conclusions}

In this paper, we proposed a sharp-interface mathematical model for simulating solid-state dewetting of thin films on a non-flat rigid substrate in two dimensions, and applied this model to studying several interesting phenomena about solid-state dewetting problems on a non-flat substrate.

First, we rigorously derived the governing equations of solid-state dewetting from the thermodynamic variation of the total interfacial free energy functional. The morphology evolution of thin films is governed by surface diffusion and contact point migration on a non-flat rigid substrate curve. Similar to the flat substrate case~\cite{Wang15,Jiang16}, we introduced a relaxation kinetics with a finite contact point mobility for describing the contact point migration. For equilibrium shapes, we obtained a bivariate equation (referred to as the generalized Young equation) to determine the static intrinsic and extrinsic contact angles of equilibrium shapes. This generalized Young equation will reduce to the classical isotropic/anisotropic Young equation when the substrate is flat~\cite{Min06,Wang15,Pierre16,Bao17b}.

Second, we used a parametric finite element method for numerically solving the proposed mathematical model.
Ample numerical experiments were performed for examining several interesting examples about solid-state dewetting
of thin films on curved substrates, i.e., equilibrium shapes of small islands, pinch-off of large islands, migration of ``small'' solid particles on curved substrates and template-assisted solid-state dewetting on a pre-patterned sinusoidal substrate. For equilibrium shapes of small islands, we found that on curved substrates different initial shapes may evolve into different equilibrium morphologies, even for the isotropic case, and the weak anisotropy also can lead to asymmetric equilibrium shapes with multiple intrinsic contact angles. For the pinch-off of large islands, we found that the critical pinch-off length $L_c$ becomes larger when the isotropic Young angle $\theta_i$ decreases and the radius $R$ of the circular substrate increases, respectively, and a simple fitting formula for $L_c$ as a function of $\theta_i$ and $R$ is also given. For a ``small'' solid particle migration on a curved substrate with a constant substrate curvature gradient $\hat{\kappa}'$, our numerical results demonstrated that the migration velocity $v$ is proportional to $\hat{\kappa}'$, inversely proportional to the square root of the area of the particle $\sqrt{A}$, and furthermore, it decreases when the isotropic Young angle increases from $0$ to $\pi$. For templated solid-state dewetting of thin films on a sinusoidal substrate, we observed four periodic categories of dewetting which have been experimentally and theoretically studied for a similar pre-patterned substrate in the reference~\cite{Giermann05}. Our simulation results are able to capture many of the complexities associated with solid-state dewetting experiments on pre-patterned curved substrates~\cite{Ahn80,Kim85,Giermann05,Giermann11}.

%
%

\section*{Acknowledgement}

This work was partially supported by the National Natural Science
Foundation of China Nos. 11871384 (W.J.) and 91630207 (Y.W. and W.B.),
Natural Science Foundation of Hubei Province No. 2018CFB466 (W.J.),
the NSF Division of Materials Research through award DMR 1609267 (D.J.S.)
and the Ministry of Education of Singapore grant R-146-000-
247-114 (W.B.). This work was partially done while the authors were visiting the Institute
for Mathematical Sciences, National University of Singapore, in 2018.

\appendix

\section{First variation to the energy functional}

In order to calculate the first variation of the total free energy functional, i.e., Eq.~\eqref{eq:energy}, we first consider an infinitesimal perturbation of the interface curve $\Gamma:= \mathbf{X}(s) = \big(x(s),y(s)\big)$, with arc
length $s \in [0,L]$, along its normal and tangent directions:
\begin{equation}
\Gamma^{\epsilon} = \Gamma + \epsilon \varphi(s)\mathbf{n} + \epsilon \psi(s)\bmath{\tau},
\label{eq:perturbed}
\end{equation}
where the perturbation parameter $\epsilon$ represents an infinitesimal number which controls the magnitude of the perturbation, and $\varphi(s),\psi(s)$ are smooth functions with respect to arc length $s$. Then the two components of the new curve $\Gamma^{\epsilon}$ can be expressed as follows:
\begin{equation*}
\Gamma^{\epsilon} = \mathbf{X}(s) + \epsilon \, \bmath{\vartheta}(s),
\end{equation*}
where $\bmath{\vartheta}(s):=(u(s), v(s))$ represents an increment vector (which is related with the direction of the position increment), and from Eq.~\eqref{eq:perturbed}, its two components along the $x$-axis and $y$-axis are easily obtained as
\begin{equation}\label{eq:def_uv}
    \begin{cases}
        u(s)=-y_s(s)\varphi(s) + x_s(s)\psi(s),\\[0.6em]
        v(s)=x_s(s)\varphi(s)+y_s(s)\psi(s).
    \end{cases}
\end{equation}
Equivalently, the function $\varphi(s)$ and $\psi(s)$ can also be expressed as:
\begin{equation}\label{eq:pp}
    \begin{cases}
        \varphi (s) = x_s(s) v(s) - y_s(s) u(s) = \bmath{\vartheta}(s) \cdot \mathbf{n}(s),\\[0.6em]
        \psi (s) = x_s(s) u(s) + y_s(s) v(s) = \bmath{\vartheta}(s) \cdot \bmath{\tau}(s).
    \end{cases}
\end{equation}
Because the contact points must move along the curved rigid substrate, the increment vectors at the two contact points must be parallel to the unit tangent vectors of substrate curve $\Gammasub$, i.e.,
\begin{equation}\label{eq:vel_c}
        \bmath{\vartheta}(0) = \lambda_l\,\Tsub(c_l),\quad
        \bmath{\vartheta}(L) = \lambda_r\,\Tsub(c_r),
\end{equation}
where $\lambda_r, \lambda_l$ are the magnitude of the increment vectors.

Therefore, the total free energy $W^{\epsilon}$ of the system with respect to
the new curve $\Gamma^{\epsilon}$ can be calculated as follows:
\begin{eqnarray}\label{eq:energy_p}
&&W^{\epsilon}\nonumber \\
&=&\int_{\Gamma^{\epsilon}}\gamma(\theta^{\epsilon})\;d\Gamma^{\epsilon}+\big(\gamma_{\subFS}-\gamma_{\subVS}\big) \Big[\big(c_r+\epsilon \lambda_r \big) -\big(c_l+\epsilon \lambda_l\big)\Big]  \nonumber  \\
&=&\int_0^L \gamma(\theta^{\epsilon})\sqrt{(x_s+\epsilon u_s)^2+(y_s+\epsilon v_s)^2}\;ds  \nonumber  \\
&&+~\big(\gamma_{\subFS}-\gamma_{\subVS}\big)\Big[\big(c_r+\epsilon \lambda_r \big) -\big(c_l+\epsilon \lambda_l\big)\Big],
\end{eqnarray}
where $\theta^\epsilon \in [-\pi, \pi]$ can be defined as the following generalization of the arctangent function:
\begin{eqnarray*}\label{eq:arctangent}
\theta^\epsilon&=&\text{Arctan}\big(\frac{y_s^\epsilon}{x_s^\epsilon}\big) \nonumber \\
&:=&\left\{
\begin{array}{ll}
\displaystyle \arctan\frac{y_s^\epsilon}{x_s^\epsilon}, & x^\epsilon_s>0, \\[0.8em]
\displaystyle \arctan\frac{y_s^\epsilon}{x_s^\epsilon}+\pi, &x^\epsilon_s<0, y^\epsilon_s \geq 0, \\[0.8em]
\displaystyle \arctan\frac{y_s^\epsilon}{x_s^\epsilon}-\pi, &x^\epsilon_s<0, y^\epsilon_s < 0, \\[0.8em]
\frac{\pi}{2}, &x^\epsilon_s=0, y^\epsilon_s > 0, \\[0.8em]
-\frac{\pi}{2}, &x^\epsilon_s=0, y^\epsilon_s < 0, \\[0.8em]
0, &x^\epsilon_s=0, y^\epsilon_s=0,
\end{array}
\right.
\end{eqnarray*}
where $x^\epsilon_s = x_s +\epsilon u_s$ and $y^\epsilon_s = y_s +\epsilon v_s$.

Then, inserting Eq.~\eqref{eq:def_uv} into Eq.~\eqref{eq:energy_p}, we can calculate its energy change rate about the curve $\Gamma$  because of this infinitesimal perturbation with respect to $\epsilon$:
\begin{eqnarray}\label{eq:variation1}
&&\frac{dW^{\epsilon}}{d\epsilon}\Big |_{\epsilon=0}=
\lim_{\epsilon\rightarrow 0}\frac{W^{\epsilon}-W}{\epsilon}\nonumber \\
&=& \int_0^L\Big(\gamma\,'(\theta) (v_s x_s - y_s u_s) + \gamma(\theta) (x_s u_s + y_s v_s)\Big)\,ds\nonumber\\
&&  +~\big(\gamma_{\subFS}-\gamma_{\subVS}\big)(\lambda_r - \lambda_l) \nonumber \\
&=& \int_0^L \Big[\gamma\,'(\theta)(\varphi_s-\kappa\psi)+\gamma(\theta) (\kappa\varphi+\psi_s)\Big)\;ds\nonumber\\
&& +~\big(\gamma_{\subFS}-\gamma_{\subVS}\big)(\lambda_r - \lambda_l)  \nonumber \\
&=& \int_0^L \big(\gamma(\theta) + \gamma\,''(\theta)\big)\kappa\varphi\;ds\nonumber \\
&&+~\Big(\gamma\,'(\theta)\varphi+\gamma(\theta) \psi+\big(\gamma_{\subFS}-\gamma_{\subVS}\big)\lambda_r\Big)_{s=L}  \nonumber \\
&&-~\Big(\gamma\,'(\theta)\varphi+\gamma(\theta)\psi+\big(\gamma_{\subFS}-\gamma_{\subVS}\big)\lambda_l\Big) _{s=0},
\end{eqnarray}
where $\kappa=-y_{ss}x_s+x_{ss}y_s$ is the curvature of the curve.

Since the two contact points must move along the curved substrate, we can obtain the following relations for $\varphi, \psi$ at $s=0$ and $s=L$ by combining the above Eq.~\eqref{eq:pp} and Eq.~\eqref{eq:vel_c}:
\begin{subequations}\label{eq:phi_psi}
    \begin{equation}
        \varphi(0) = \lambda_l\; \Tsub(c_l) \cdot \mathbf{n}(0) = -\lambda_l \sin\theta_{\rm i}^l,
    \end{equation}
    \begin{equation}
        \psi(0) = \lambda_l \;\Tsub(c_l) \cdot \bmath{\tau}(0) = \lambda_l \cos\theta_{\rm i}^l,
    \end{equation}
    \begin{equation}
        \varphi(L) = \lambda_r \;\Tsub(c_r) \cdot \mathbf{n}(L) = -\lambda_r \sin\theta_{\rm i}^r,
    \end{equation}
    \begin{equation}
        \psi(L) = \lambda_r \;\Tsub(c_r) \cdot \bmath{\tau}(L) = \lambda_r \cos\theta_{\rm i}^r.
    \end{equation}
\end{subequations}
Therefore, Eq.~\eqref{eq:variation1} can be rewritten as follows
\begin{eqnarray}\label{eq:variation}
    &&\frac{dW^{\epsilon}}{d\epsilon}\Big |_{\epsilon=0} = \int_0^L \big(\gamma(\theta) + \gamma\,''(\theta)\big)\kappa\varphi\;ds \nonumber\\
    &&\quad +\Big[\gamma(\theta_{\rm e}^r) \cos\theta_{\rm i}^r - \gamma\,'(\theta_{\rm e}^r)\sin\theta_{\rm i}^r + \big(\gamma_{\subFS}-\gamma_{\subVS}\big)\Big]\lambda_r \nonumber\\
    &&\quad -\Big[\gamma(\theta_{\rm e}^l) \cos\theta_{\rm i}^l - \gamma\,'(\theta_{\rm e}^l)\sin\theta_{\rm i}^l + \big(\gamma_{\subFS}-\gamma_{\subVS}\big)\Big]\lambda_l.\nonumber
\end{eqnarray}

\section{Mass conservation and energy dissipation}

We introduce a new variable $p\in I=[0,1]$, which is independent of time $t$, to parameterize the moving film/vapor interface as $\Gamma(t)=\vec X(p,t)=(x(p,t),y(p,t))$, where $p=0$ and $p=1$ are used to represent the left and right contact points, respectively.  The relationship between the parameter $p$ and the arc length $s$ can be given as $s(p,t)=\int_0^p |\partial_p\vec{X}|\;dp$, and then we can obtain that $\partial_p s=|\partial_p\vec{X}|$. For simplicity, we use subscripts to denote partial derivatives, such as $s_p = \partial_p s, \mathbf{X}_p = \partial_p \mathbf{X}$.

\begin{proof}[Proof of mass conservation]

The dimensionless total area (or mass) of the thin film on the curved substrate $\hat{\Gamma}:=(\hat{x}(c),\hat{y}(c))$
is defined as
\begin{equation*}
  A(t) = \int_0^1 y x_p \,dp - \int_{c_l}^{c_r} \hat{y} \hat{x}_c \, dc.
\end{equation*}
Therefore, the rate of change of the dimensionless total area (or mass) can be calculated as
\begin{eqnarray}
    \frac{d A}{d t} & = & \int_0^1 (y_t x_p + y x_{pt})\,dp - \Big(\hat{y}\hat{x}_c\frac{dc_r}{dt}\Big)_{c = c_r}+ \Big(\hat{y}\hat{x}_c\frac{dc_l}{dt}\Big)_{c = c_l} \nonumber \\
    & = & \int_0^1 (y_t x_p - y_p x_t)\,dp + y x_t\Big|_{p=0}^{p=1}\nonumber\\
    && -~ \hat{y}(c_r) \frac{d \hat{x}(c_r)}{dt} + \hat{y}(c_l) \frac{d \hat{x}(c_l)}{dt}\nonumber\\
    & = &\int_0^1 (x_t, y_t)\cdot(-y_p, x_p) \,dp \nonumber\\
    & = &\int_0^{L(t)} \mathbf{X}_t\cdot\mathbf{n} \,ds = \int_0^{L(t)} \mu_{ss} \,ds\nonumber\\
    & = &\mu_s\big(L(t),t\big) - \mu_s\big(0,t\big) = 0\nonumber
\end{eqnarray}
During the above calculations, we have used the integration by parts, and that the contact points simultaneously lie on the film/vapor interface $\Gamma$ and the substrate curve $\hat{\Gamma}$, i.e.,
\begin{gather*}
    \big(x(0,t), \quad y(0,t)\big) = \big(\hat{x}(c_l), \quad \hat{y}(c_l)\big),\\
    \big(x(1,t), \quad y(1,t)\big) = \big(\hat{x}(c_r), \quad \hat{y}(c_r)\big).
\end{gather*}

Hence, by using the zero-mass flux condition $\mu_s (0, t) = \mu_s (L, t) = 0$,
the total area (mass) of the thin film is conserved during the evolution.
\end{proof}

\begin{proof}[Proof of energy dissipation]

The dimensionless total free energy of the system defined in Eq.~\eqref{eq:energy} can be rewritten as:
\begin{equation*}
    W(t) = \int_0^1 \gamma(\theta)s_p \;dp - \sigma \big(c_r(t)-c_l(t)\big),
\end{equation*}
where $\sigma := (\gamma_{\subVS} - \gamma_{\subFS})/\gamma_0$. Following with the similar method we used in the flat substrate case~\cite{Wang15}, we can calculate the rate of change of the dimensionless total free energy as follows:
\begin{eqnarray}\label{eq:DW}
    \frac{d W}{d t} & = & \int_0^1 \Big(\gamma\,'(\theta) \theta_t s_p + \gamma(\theta) s_{pt}\Big)\,dp
    -\sigma\Big(\frac{d c_r}{dt} - \frac{d c_l}{dt}\Big) \nonumber \\
    & = & \int_0^1 \mathbf{X}_{pt}\cdot \big(\gamma\,'(\theta)\,\mathbf{n} + \gamma(\theta)\, \bmath{\tau}\big)\,dp -\sigma\Big(\frac{d c_r}{dt} - \frac{d c_l}{dt}\Big) \nonumber \\
    & = & - \int_0^1 \mathbf{X}_t\cdot \Big(\big(\gamma\,''(\theta)\theta_p \mathbf{n} + \gamma\,'(\theta) \kappa s_p \bmath{\tau}\big) \nonumber\\
    &&+~ \big(\gamma\,'(\theta) \theta_p \bmath{\tau} - \gamma(\theta)\kappa s_p \mathbf{n}\Big)\,dp \nonumber\\
    && +~ \Big(\mathbf{X}_{t}\cdot \big(\gamma\,'(\theta)\,\mathbf{n} + \gamma(\theta)\, \bmath{\tau}\big)\Big)_{p=0}^{p=1} \nonumber\\
    &&-\sigma\Big(\frac{d c_r}{dt} - \frac{d c_l}{dt}\Big) \nonumber\\
    & = & \int_0^{L(t)} \kappa\Big(\gamma(\theta)+\gamma\,''(\theta)\Big)\, \mathbf{X}_t\cdot\mathbf{n} \,ds\nonumber\\
    && +~ \frac{d c_r}{dt}\Big(\gamma(\theta_{\rm e}^r) \cos\theta_{\rm i}^r - \gamma\,'(\theta_{\rm e}^r) \sin\theta_{\rm i}^r -\sigma\Big)\nonumber \\
    &&-~ \frac{d c_l}{dt}\Big(\gamma(\theta_{\rm e}^l) \cos\theta_{\rm i}^l - \gamma\,'(\theta_{\rm e}^l) \sin\theta_{\rm i}^l -\sigma\Big)\nonumber \\
    & = &\int_0^{L(t)} \mu \mu_{ss} \,ds - C\bigg[\Big(\frac{d c_r}{dt}\Big)^2 + \Big(\frac{d c_l}{dt}\Big)^2\bigg]\nonumber \\
    & = &\mu \mu_s\Big|_{s=0}^{s=L(t)} - \int_0^{L(t)} \mu_s^2\,ds - C\bigg[\Big(\frac{d c_r}{dt}\Big)^2 + \Big(\frac{d c_l}{dt}\Big)^2\bigg]\nonumber \\
    & = & - \int_0^{L(t)} \mu_s^2\,ds - C\bigg[\Big(\frac{d c_r}{dt}\Big)^2 + \Big(\frac{d c_l}{dt}\Big)^2\bigg] < 0, \nonumber
\end{eqnarray}
where the constant $C=1/\eta>0$ for $0<\eta<\infty$. In the above calculations, we have used integration by parts, the relaxed contact angle boundary conditions and the zero-mass flux condition.

Hence, the total free energy of the system decreases during the evolution.
\end{proof}

\section{Numerical algorithm}

We implement the proposed sharp-interface model by a semi-implicit parametric finite element method~\cite{Bao17}. In this appendix, we briefly present its variational formulation and the corresponding finite element approximation.

\

\noindent \emph{1. Variational formulation}

Given an initial curve $\Gamma(0)=\vec X(p,0), p \in I=[0,1]$, for $t \in (0,T]$, find the evolution curves $\Gamma(t)=\vec X(p,t)\in H_{a,b}^1(I)\times H_{c,d}^{1}(I)$, the chemical potential $\mu(p,t)\in H^1(I)$, and the curvature $\kappa(p,t)\in H^1(I)$ such that
\begin{eqnarray}
&&\big<\vec{X}_t,~\varphi\mathbf{n}\big>+\big<\mu_s,~\varphi_s\big>=0, \forall \varphi\in H^{1}(I), \label{eqn:weq1}\\[1.0em]
&&\big<\mu,~\phi\big>-\big<\widetilde{\gamma}(\theta)\kappa,~\phi\big>=0, \forall \phi\in H^{1}(I), \label{eqn:weq2}\\[1.0em]
&&\big<\kappa\mathbf{n},~\boldsymbol{\omega}\big>-\big<\vec{X}_s,~\boldsymbol{\omega}_s\big>=0, \forall \boldsymbol{\omega}\in H_{0}^{1}(I)\times H_{0}^{1}(I), \label{eqn:weq3}
\end{eqnarray}
where $H^1(I)$ and $H^1_0(I)$ are the standard Sobolev space with the
derivative taken in the distributional or weak sense~\cite{Brenner07}, and $a, b, c, d$ stand for the $x$-coordinates and $y$-coordinates of the left and right contact points, i.e., $x(0,t), x(1,t), y(0,t), y(1,t)$ at time $t$, respectively. The functional space $H_{a, b}^1(I)$ is defined as follows
\begin{equation*}
H_{a,b}^1(I)=\{f \in H^{1}(I): f(0)=a, f(1)=b\},
\end{equation*}
and $H^1_0(I):=H^1_{0,0}(I)$. The symbol $\big<\cdot,~\cdot\big>$ is the $L^2$ inner product with respect to the curve $\Gamma(t)$ defined as follows
$$\big<f,g\big>:=\int_{\Gamma(t)}f\cdot g\;ds,$$
where $f, g$ are scalar (or vector) functions.

In fact, the above variational formulation \eqref{eqn:weq1} is obtained by reformulating the first equation in Eqs.~\eqref{eq:GE_curve_weak} as
$\vec{X}_t\cdot \vec{n}=\mu_{ss}$, multiplying the test function
$\varphi$, integrating over $\Gamma$, integration by parts and using
the boundary condition \eqref{eq:BC3_curve_weak}. Similarly, \eqref{eqn:weq2} is derived
from $\mu=\widetilde{\gamma}(\theta)\kappa$ by multiplying the test function
$\phi$, and \eqref{eqn:weq3} is obtained
from the second equation in Eqs.~\eqref{eq:GE_curve_weak} by reformulating it as
$\kappa\vec{n}=-\partial_{ss}\vec{X}$ and taking the dot-product with the test function
$\boldsymbol{\omega}$. For more details, the readers can refer to \cite{Bao17}.

\

\noindent \emph{2. Finite element approximation}

First, we decompose $I$ into $N$ small intervals
$$I=[0,1]=\bigcup_{j=1}^{N}I_j=\bigcup_{j=1}^{N}[q_{j-1},q_{j}],$$ with the nodes $q_j = jh, h = 1/N$.
In addition, let $0=t_{0}<t_{1}<\ldots<t_{M-1}<t_{M}=T$ be a partitioning of the time interval $[0,T]$.
Define $\Gamma^{m}=\vec{X}^{m}$ as the numerical approximation to the moving curve $\Gamma(t_m)$. Similarly, we can define other numerical approximation notations, e.g.,
$\mathbf{n}^m, \mu^m, \kappa^m$. Then, we define the conforming finite element spaces for the numerical approximation solution as follows:
\begin{equation}
\label{eqn:FEMspace1}
V^h:=\{u\in C(I): u\mid_{I_{j}}\in P_1,\; \forall \, j=1,2,\ldots,M\}
\end{equation}
\begin{equation}
\label{eqn:FEMspace2}
\mathcal{V}^h_{a,b}:=\{u \in V^h:\;u(0)=a,~u(1)=b\},
\end{equation}
where $a$ and $b$ are two given parameters related with the two moving contact points, and for simplicity we
denote the solution space $\mathcal{V}^h_0=\mathcal{V}^h_{0,0}$. The normal vector of the numerical solution $\Gamma^m$, which is a step function with possible discontinuities or jumps at nodes $q_j$, can be computed as $$\mathbf{n}^m=-\bigl(\vec {X}_s^m\bigr)^{\perp}=-\frac{\bigl(\vec{X}_p^m\bigr)^{\perp}}{|\vec{X}_p^m|},$$
where ``$\perp$'' denotes a clockwise rotation through $90$ degrees.

For any two scalar (or vector) functions $u$ and $v$, we define the $L^2$ inner product $\big<u,v\big>_{\Gamma^m}$
over the current polygonal curve $\Gamma^m$ at the time level $t^m$ as follows:
\begin{equation*}
\big<u,v\big>_{\Gamma^m}=\int_{\Gamma^m}(u \cdot v)\;ds=\int_{0}^{1}(u\cdot v)|\vec{X}_p^m|\;dp.
\end{equation*}
Furthermore, if $u$ and $v$ are two piecewise continuous scalar or vector functions defined on the domain $I$, with possible jumps at the nodes $\{q_j\}_{j=1}^N$, we can define the mass lumped inner product $\big<\cdot,\cdot\big>_{\Gamma^m}^h$ as follows:
\begin{equation*}
\big<u,~v\big>_{\Gamma^m}^h=\frac{h}{2}\sum_{j=1}^{N}\Big|\vec{X}_p^m(q_{j-\frac{1}{2}})\Big|\Big[\big(u\cdot v\big)(q_j^-)+\big(u\cdot v\big)(q_{j-1}^+)\Big],
\end{equation*}
Where $u(q_j^-)$ and $u(q_j^+)$ represent the limit values at the the possible jump
node $q_j$ from the left-hand side and the right-hand side, respectively.

The parametric finite element approximation to the weak formulation~\eqref{eqn:weq1}-\eqref{eqn:weq3} can be described as follows:

Given the curve $\Gamma^m=\vec{X}^m$ at the time level $t_m$, for the next time level $t_{m+1}$, find the evolution curve $\Gamma^{m+1}=\vec{X}^{m+1}\in \mathcal{V}^h_{a,b}\times \mathcal{V}^h_{c, d}$ with $a, b$ the $x$-coordinates of the two contact points at $t_{m+1}$ and $c, d$ the $y$-coordinates, the chemical potential $\mu^{m+1}\in V^h$ and the curvature $\kappa^{m+1}\in V^h$ such that:
\begin{gather*}
\Big<\frac{\vec{X}^{m+1}-\vec X^{m}}{t_{m+1} - t_m},\varphi\mathbf{n}^{m}\Big>_{\Gamma^{m}}^{h}+\big<\mu_s^{m+1},\varphi_s\big>_{\Gamma^{m}}=0,
\,\forall~\varphi\in V^h,     \nonumber \\[1.0em]
\big<\mu^{m+1},~\phi\big>_{\Gamma^{m}}^{h}-\big<\widetilde{\gamma}(\theta^m)\kappa^{m+1},
~\phi\big>_{\Gamma^{m}}^{h}=0,
\,\forall~\phi\in V^h, \nonumber\\[1.0em]
\big<\kappa^{m+1}\mathbf{n}^{m},~\boldsymbol{\omega}\big>_{\Gamma^{m}}^{h}-\big<\vec {X}_s^{m+1},~\boldsymbol{\omega}_s\big>_{\Gamma^{m}}=0,
\,\forall~\boldsymbol{\omega}\in \mathcal{V}^h_0\times \mathcal{V}^h_0, \nonumber
\end{gather*}
where the arc lengths of moving contact points, i.e. $c_l(t_{m+1})$ and $c_r(t_{m+1})$, are updated by solving the relaxed contact angle condition, Eq.~\eqref{eq:BC2_curve_weak}, via the forward Euler scheme. Then, according to the values of
$c_l(t_{m+1})$ and $c_r(t_{m+1})$, we can obtain the values of the parameters $a,b,c,d$ at the time level $t_{m+1}$ by using
the formula of substrate curve.

\bibliography{mybib}

\end{document}